# On the degeneracy of solutions from stellar spectropolarimetric data

J. C. Ramírez Vélez,[1]* J. M. Raygoza-Romero,[2] I. H. Lopez-Nava,[2] R. López -Valdivia,[1]
M. F. Arenas[1] and L. Adame Villanueva[1]
[1] *Universidad Nacional Autónoma de México, Instituto de Astronomía, A.P. 106, Ensenada 22800, B.C., México*
[2]*Computer Science Department, Centro de Investigación Científica y Educación Superior de Ensenada, C. Tijuana-Ensenada 3918, Ensenada 22860, México*

**ABSTRACT**

In this work we investigate the inversion accuracy of stellar magnetic fields through the analysis of spectropolarimetric data. We performed several noise-free tests to untangle the impact among the atmospheric and magnetic parameters in the data analysis. Using a single spectral line, we found that under ideal scenario, without noise, the magnetic parameters can be recovered with very high accuracy, while some of the atmospheric ones show a considerable incertitude. If multi-line profiles are considered instead, then the accuracy of the recovered atmospheric parameters increases significantly. Nonetheless, it is quite common that for the inversion of stellar spectropolarimetric data, the atmospheric parameters and the stellar inclination angle of the star are fixed; we show that this procedure could induce very high errors in the inference of the magnetic properties, specially in what concerns to fix the inclination angle. We show that even small deviations from the true inclination angles could have as consequence that the recovery of the magnetic properties of the star are no longer reliable. This is because we demonstrate that given a set of Stokes profiles observed along the rotational phase, they can be satisfactorily fitted regardless of the assumed stellar inclination angle, thereby revealing a degeneracy in the solution. Although this work validates the solution generation solely for a de-centered dipolar geometry, a comparable degeneracy is anticipated when modeling the magnetic field with higher-order spherical harmonics—an effect that remains to be formally characterized.

## 1 INTRODUCTION

Spectropolarimetry has established itself as the most robust technique for mapping stellar magnetic fields, enabling the reconstruction of field strength and topology through the inversion of Stokes profiles. However, the inversion of these profiles represents a major computational and physical challenge, characterized as a highly non-linear and degenerate inverse problem. In this context, magnetic parameters—such as the magnetic topology, inclination ($i$) of the rotation axe respect to the line-of-sigth (LOS), and geometric orientation—are intimately coupled with the rest of parameters, including effective temperature ($T_{\rm eff}$), surface gravity (log g), metallicity ([M/H]), and projected rotational velocity (vsini), which significantly complicates their accurate retrieval.

In solar physics, the development of inversion codes has attained a high degree of technical sophistication through rigorous validation protocols. Pioneering tools such as SIR (Ruiz Cobo & del Toro Iniesta 1992) introduced the use of response functions to retrieve the stratification of physical parameters. These are validated via stability tests where synthetic observations, generated from known solar models, are inverted using diverse random initial atmospheres; such tests demonstrate convergence by aligning results with the reference model with minimal root-mean-squared (RMS) deviations, independent of initial conditions. Other advanced codes, such as HAZEL (Asensio Ramos et al. 2008), validate the resolution of ambiguities—such as the Van Vleck effect—by employing the DIRECT (DIviding RECTangles) global optimization method (Jones et al. 1993) for initialization and the Levenberg-Marquardt algorithm (Press et al. 1986) for refinement, achieving precise agreement with solar prominence observations and theoretical models of atomic polarization.

NICOLE (Socas-Navarro et al. 2015) and STiC (de la Cruz Rodríguez et al. 2019) not only employ massive parallelization and 3D magnetohydrodynamic (MHD) models under non-local thermodynamic equilibrium (non-LTE) conditions to validate spectral profile inversions, but they also implement two methods for the formal solution of the radiative transfer equation (RTE): one based on a parabolic interpolation and another utilizing splines functions, which control overshooting while maintaining accuracy. Recently, the DeSiRe code (Ruiz Cobo et al. 2022) has integrated these capabilities for the study of complex solar layers. However, the extreme computational cost of these tools—requiring supercomputers in the case of NICOLE—limits their application in systematic explorations of disk-integrated stars, where the geometry is inherently unknown.

Furthermore, within stellar astrophysics, widely adopted codes such as INVERS10/13 (Piskunov & Kochukhov 2002; Kochukhov & Piskunov 2002; Kochukhov et al. 2013) originally spearheaded the magnetic surface mapping of chemically peculiar (Ap/Bp) stars, before to extend the study sample to active RS CVn-type systems, including II Pegasi. Their validation protocol is through synthetic Stokes profiles generated for stars with topologies ranging from simple dipoles to complex multipolar configurations; these profiles are then convolved with instrumental profiles and affected with Gaussian noise to reach profiles with a signal-to-noise ratio (S/N) of about 300; inversions are performed using a Levenberg-Marquardt algorithm considering over at least 10 rotational phases, and the code employs a Tikhonov regularsion method. The accuracy of the test is assessed by comparing recovered maps with the originals one through mean errors in field strength and orientation, expanding the results into spherical harmonics to verify the retrieval of high-order components. These tests have show that the code is robust against moderate errors in inclination and vsini. However, these approaches typically assume or fix most of the stellar atmospheric parameters, coupled with the subjacent fact that the code was tested for the non-magnetic case and is currently being applied to magnetic scenarios (Kochukhov 2017). Additionally, the reliability of some results derived from these codes remains under debate. Specifically, the produced magnetic and elemental abundance maps for chemically peculiar stars seem to exhibit inconsistencies, even when based on the same datasets (Stift & Leone 2026).

* E-mail: jramirez@astro.unam.mx

Concurrently, iMap (Carroll et al. 2007, 2012) and its extension OMP (Carroll & Strassmeier 2014) utilize conjugate gradient techniques and sparse representations to investigate active late-type stars, such as II Pegasi, and pre-main sequence objects like V410 Tau. These methods validate their performance through hundreds of inversions with randomized initial conditions—attaining a typical dispersion of $\sim$ 64 G in the radial field component—as well as through stellar dynamo simulations with added noise. In this context, they demonstrate remarkable robustness against noise, handling residual errors in the Stokes profiles on the order of $10^{-3}$ to $10^{-5}$ without inducing spurious solutions, and achieving global accuracies of $\sim$ 2–5%, thereby outperforming traditional methods under low S/N conditions. However, the authors do not explore magneto-atmospheric degeneracies.

Furthermore, codes such as PIMMS (Gutteridge et al. 2026) and TIMES (Finociety & Donati 2022) have extended these approaches into the time domain, incorporating stellar variability and pulsations in objects such as $\beta$ Cep and SPB stars through maximum entropy regularization principles (Brown et al. 1991; Donati & Brown 1997; Donati et al. 2006). In particular, PIMMS is notable for its tests with high-quality synthetic data (S/N $\approx$ 1900), using models with $\sim$ 10,000 surface cells to reproduce pulsation-induced brightness and magnetic field variations, as in the case of the $\beta$ Cep star V2052 Oph. This approach successfully retrieves brightness distributions, including isolated polar spots, with errors below 2%, where traditional methods fail by not accounting for time-dependent surface velocities. However, it requires between 19 and 109 spectra per rotational cycle to converge. Complementarily, TIMES validates the reconstruction of poloidal and toroidal topologies under high observational cadence conditions, achieving $\chi^2 \sim 1.02$ values, in contrast to the $\chi^2 \sim 8.5$ obtained by conventional methods. Despite these advancements, both approaches continue to assume as known the atmospheric parameters and the inclination angle of the star.

Despite these advancements, in the stellar domain a critical limitation persists: most of these methods fix the atmospheric parameters ($T_{\rm eff}$, log g, vsini, and inclination $i$) derived from unpolarized spectra to maintain computational feasibility. Gradient-based or direct-synthesis approaches become prohibitively expensive or unstable as the dimensionality of the parameter space increases, preventing a systematic exploration of global magneto-atmospheric degeneracies. Consequently, it remains unquantified how uncertainties in parameters such as temperature, metallicity or inclination propagate into the magnetic field inference.

In this work, we present MAPNet, which overcomes these limitations by simultaneously treating 12 magneto-atmospheric parameters as free parameters, enabling for the first time systematic tests of full degeneracies through massive, GPU-accelerated inversions.

## 2 STOKES PROFILE SYNTHESIS

In this study, the synthesis of the Stokes profiles will be carried out using the polarized radiative transfer code COSSAM[1] code (Stift et al. 2012). However, the use of the code is not straightforward but it is done by means of an Artificial Neuronal Network (ANN). The basis of this approach as well as the performance of the ANN are described in detail in Raygoza-Romero et al. (2025), hereafter Paper I.

While in Paper I, for the profile synthesis we only considered 8 magnetic parameters required to describe the de-centered oblique rotator model (e.g. Stift 1975), in the present work we additionally incorporated 4 atmospheric parameters. Therefore, as a first step, we will start showing that this new ANN is able to synthesizes the Stokes profiles very closely as the COSSAM code does: this means considering 12 parameters, 8 for the magnetic model and 4 for the atmospheric one. We have named MAPNet this new ANN as it stands for *M*agnetic and *A*tmospheric *P*arameters with neuronal *Net*works.

Typically, in the analysis of stellar spectropolarimetric data, excepting in some rare cases of Ap/Bp stars, the atmospheric parameters such as effective temperature, gravity and metallicity are determined a priori to the inversion process, but here we have allowed to vary them during the fitting process in order to inspect if exist any type of solution degeneration between magnetic and atmospheric parameters.

It is important to remember that most of the observational studies devoted to magnetic stars usually select targets where at least $T_{\rm eff}$ and log g are known, and for this reason we decided to limit to $\pm 250$ K for $T_{\rm eff}$ and $\pm 0.5$ dex for log g, around some values that could be found in literature. In our case, we assume an hypothetical star with $T_{\rm eff}$ = 7250 K and log g = 4.0 dex.

The input of MAPNet are the magneto-atmospheric model parameters and the output are the four Stokes profiles (I,Q,U,V). The range variations of each parameter are shown in Table 1, where $m$ is the magnetic moment of the dipole, $i$ is the inclination angle of the rotation axis of the star respect to the line-of-sight (LOS), $\alpha$, $\beta$, and $\gamma$ are the eulerian angles, $x_2$ and $x_3$ determine the position of the dipole inside of the star, $\phi$ denotes the phase rotation of the star, and, vsini corresponds to the rotation velocity of the star protected along the LOS. An important parameter that is not included in Table 1 will be the position of the magnetic dipole inside the star given by $x_r = \sqrt{(x_2^2 + x_3^2)}$.

MAPNet is an ANN with 7 hidden fully connected layers each one with 4,096 neurons, an input layer with 12 neurons (one per each magneto-atmospheric parameter) and an output layer with $4 \times n$ neurons, where $n$ corresponds to the number of wavelength points considered in the profile synthesis and the number 4 is due to each of the Stokes profiles (see Fig. 1 of Paper I).

For the establishment of the training database we have chose the FeI line at 4982.4 Å and we considered the highest spectral resolution achievable nowadays (R ~ 115,000) with the HARPSPol spectropolarimeter (Piskunov et al. 2011). Given the range variation values of Table 1, we used COSSAM to synthesize the four Stokes profiles considering 1.5 million of different combinations of parameter values. The line list for each atmospheric model were obtained from the VALD database (Ryabchikova et al. 2015), while the atmospheric models required by COSSAM to perform the profiles synthesis were obtained from the Kurucz-Castelli grid[2] (Castelli & Kurucz 2003), and that is the reason why $T_{eff}$, log g and [M/H] varies in steps, contrary to the rest of parameters that do it randomly (last column in Table 1). Please note that the total number of atmospheric models used from the Kurucz-Castelli grid are 63, resulting from the different combinations of $T_{eff}$, log g and [M/H] (see Table 1).

[1] https://www.ada2012.eu/cossam_ff/index.html

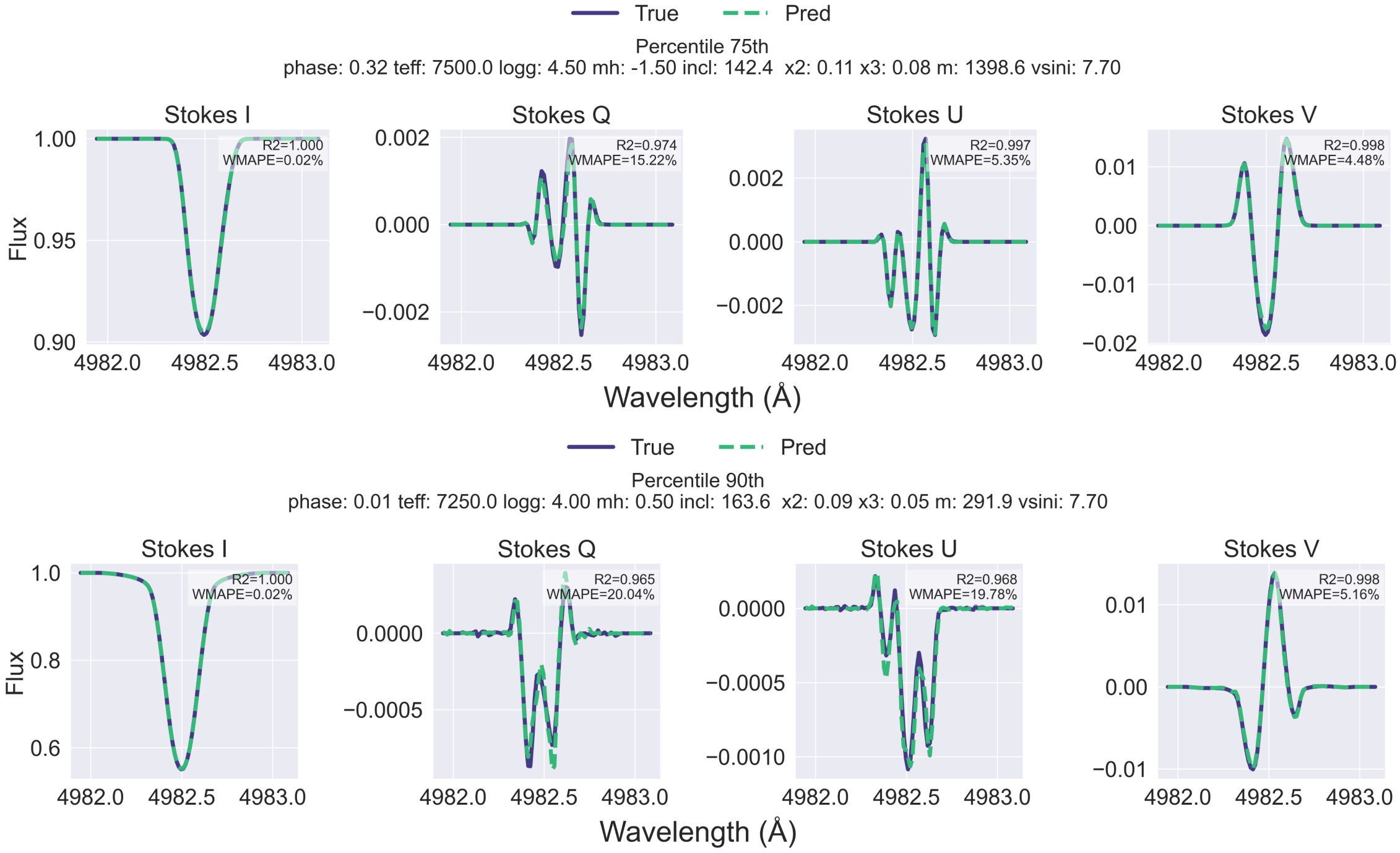

**Figure 1.** Comparison of the Stokes profiles synthesized with COSSAM (black profiles) and those by the ANN (green profiles). In the upper and lower panels are shown the 75th and 90th percentiles, meaning that the match of profiles between COSSAM and MAPNet is better in 75% and 90% of cases, respectively.

**Table 1.** Range variation of the parameters.

| | Min | Max | Variation |
|---|---|---|---|
| Magnetic | | | |
| m (Gauss) | 100 | 5100 | random |
| i (deg) | 0 | 180 | random |
| $\alpha$ , $\gamma$ (deg) | -180 | 180 | random |
| $\beta$ (deg) | 0 | 180 | random |
| $x_2$ , $x_3$ ($R_\star$) | 0 | 0.2 | random |
| $\phi$ (phase) | 0 | 1 | random |
| Atmospheric | | | |
| $T_{eff}$ (K) | 7000 | 7500 | step 250 |
| log g (dex) | 3.5 | 4.5 | step 0.5 |
| M/H (dex) | -2.0 | 0.5 | step 0.5 (+0.2) |
| vsini (km/s) | 2 | 10 | random |

The training process of MAPNet followed the same methodology as the one described in Paper I, which means, to perform multiple tests to find the best model among different ANN architectures and different scaling algorithms applied to the Stokes profiles prior to the training – this last step is known as data pre-processing (e.g. Pedregosa et al. 2011)–. In particular, the database conformed with 1.5 million instances was splitted as customary, 75% for training, 15% for validation, and 10% for testing.

Once trained, we found that the accuracy of MAPNet to synthesize the Stokes profiles is comparable to the one from the radiative transfer code. The model achieved median weighted mean absolute percentile error (WMAPE) values of (0.03, 4.7, 6.8, 2.9) for Stokes (I,Q,U,V) respectively. These results demonstrate that MAPNet is capable of accurately reproducing the polarized spectral signatures generated by the radiative transfer calculations.

In Fig.1 are shown the percentiles 75 and 90 -upper and lower panels respectively-, illustrating the very good agreement of the ANN (green profiles) respect to the COSSAM code (dark blue profiles). In each row, the value of the magneto-atmospheric parameters used in COSSAM for the profile synthesis are indicated in the subtitle. In each subplot, are indicated 2 different metrics to evaluate the errors among the profiles : the $R^2$ coefficient and the WMAPE.

An important characteristic of MAPNet is that the performance is a function of the amplitude of Stokes profiles, meaning that the lower the amplitude the higher the error in the MAPNet profiles. This is mainly important for linear Stokes profiles (Q,U); see appendix A. This behavior has been previously found in Paper I, however in the present study even for the worst cases with the smallest amplitudes in the linear profiles between $10^{-4}-10^{-3}$, the performance of MAPNet is acceptable: the $R^2$ value in the 90th percentile in Fig. 1 for the linear profiles Q and U is close to 0.97, while for profiles I and V the errors are negligible ($R^2$ close to unity). For a deeper discussion about errors as function of profiles amplitude see Paper I. Please note that Fig. 1 illustrates that MAPNet reproduce the Stokes profiles extremely well most of the time, since in 75% of cases the profiles agreement between COSSAM and MAPNet is better than one showed in the upper panels.

[2] https://wwwuser.oats.inaf.it/fiorella.castelli/grids.html

Given the performance showed by MAPNet, we conclude that this ANN can synthesize the Stokes profiles very closely as the radiative transfer code does. Besides, and as it will be show below, the small differences among MAPNet and the radiative code are not significative, meaning that the ANN can safely substitute the radiative code in the data analysis, with the advantage of the extremely fast synthesis speed of MAPNet – something that will be very relevant later.

## 3 INVERSION TEST

Once it has shown that MAPNet properly synthesizes the Stokes profiles, we will employ it to invert a set of synthetic profiles. In order to simplify the interpretation of the results, we consider the hypothetical case of Stokes profiles without noise.

Let us consider seven given known phases $\phi_j$ along the rotation period of the stars, such that $j$ varies from 1 to 7, and let $\chi_i$ denotes a given combination of atmospheric and magnetic parameters. Please note that $\chi_i$ consists of 11 values because the known phases are given by $\phi_j$. At each of the 7 phases we used COSSAM to synthesize the four Stokes parameters, such that this set of 7x4 Stokes profiles would correspond to one "observed" star. The size of the full sample of Stokes profiles corresponds to 1,298 "observed" stars.

For the inversion process, we employed a Particle Swarm Optimization (PSO) algorithm described in Kennedy & Eberhart (1995). The optimization is performed using a population of 2,048 particles that simultaneously explore the multidimensional parameter space in search of the best-fitting solution. We selected a population-based optimization strategy because MAPNet fully benefits from the high degree of parallelization provided by artificial neural networks running on Graphics Processing Units (GPUs), enabling the evaluation of a large number of candidate solutions within a reasonable computational time.

Additionally, the PSO implementation adopts a ring-topology neighborhood structure as in Kennedy (1999), where each particle exchanges information only with a limited set of neighboring particles (64 in our case), promoting a better balance between global exploration and local exploitation of the parameter space. To ensure stability during the optimization, a damping (bounce-back) boundary condition was also applied whenever particles moved outside the predefined search-space limits, preventing nonphysical solutions while preserving the swarm dynamics.

During the optimization procedure, each particle generates candidate combinations of the free parameters whose corresponding Stokes profiles are synthesized by MAPNet and compared against the observed profiles. As a result, the inversion of a single observed star involves the evaluation of a very large number of candidate solutions throughout the optimization process, enabling an extensive exploration of the multidimensional parameter space and increasing the robustness and reliability of the inferred magnetic and stellar configuration.

More specifically, each PSO execution employs a population of 2,048 particles evolved over 50 iterations, resulting in the evaluation of 102,400 candidate parameter combinations during a single inversion run. Since each candidate solution must reproduce the full rotational modulation of the star, and our observations consist of 7 rotational phases, this corresponds to the synthesis and comparison of 716,800 complete sets of Stokes profiles per PSO run. Considering that each rotational phase contains the four Stokes parameters individually (I, Q, U, and V), a total of approximately 2.8 million individual Stokes profiles are synthesized during a single optimization process.

**Table 2.** Resume of the inversions results using FeI line at 4982.4 Å. The parameters with an $R^2$ equal or superior to 0.9 has been marked in bold.

| | $T_{eff}$ (K) | log g (dex) | [M/H] (dex) | Incl (deg) | $x_r$ ($R_\star$) | m (G) | vsini (km/s) |
|---|---|---|---|---|---|---|---|
| MAE | 85 | 0.15 | **0.06** | **0.5** | **0.004** | **17** | **0.02** |
| RMSE | 134 | 0.23 | **0.08** | **0.8** | **0.009** | **24** | **0.03** |
| $R^2$ | 0.574 | 0.602 | **0.991** | **0.999** | **0.982** | **0.999** | **0.999** |

This is where the fast speed profiles synthesis of MAPNet becomes very relevant: a single PSO execution requires approximately only 9 seconds on a GPU[3]. To improve robustness against stochastic effects and reduce the probability of converging toward local minima, the inversion procedure is repeated five independent times, selecting the solution with the best fitness value. Consequently, the total computational time per observed star is approximately 45 seconds.

The inversion procedure employs the Expected Weighted Mean Absolute Percentage Error (EWMAPE) as the fitness metric to evaluate the agreement between the observed and synthesized Stokes profiles. A detailed description of the metric, together with the corresponding mathematical formulation details, is provided in Appendix A.

To illustrate the good profiles adjustment reached in the inversion, in Fig. 2 we show the percentile 75 arranged in order of their fitness, meaning that three-quarters of the sample have a similar or better fit than the one showed in this figure.

In Fig. 3 are shown the inversions results per individual parameter, while in Table 2 are resumed these results through 3 metrics: the Mean Absolute Error (MAE), the Root Mean Squared Error (RMSE) and the $R^2$ coefficient. It is surprising that only the temperature and gravity are the two parameters that cannot be recovered properly, specially in regard of the very high accuracy obtained in the rest of them. We can therefore conclude that is not possible to a priori assume that the temperature, gravity and magnetic field can all be properly inferred simultaneously from any given spectral line.. We have not inspected if this inversion incertitude in $T_{eff}$ and log g remains if another spectral line is considered in lieu of the one used here at 4982.4 Å. Instead, we have tested the inversion of multi-line profiles, which in fact correspond to what in praxis is used in the analysis of stellar spectropolarimetric data. As was mentioned before, excepting the case of spectra of intermediate and massive star with strong magnetic fields where polarization can be detected in single spectral lines, most of the studies require line-addition methods to boost the amplitude of the polarized mean profile (e.g. Semel & Li 1996; Donati et al. 1997; Martínez González et al. 2008).

### 3.1 Inversion of multi-line profiles

For the data set of multi-line profiles, we considered a spectral range from 5000 to 5100 Å. Setting a minimum line-depth criterion of 0.1

[3] In our case, we used the NVIDIA RTX A5000 model.

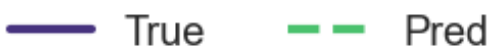

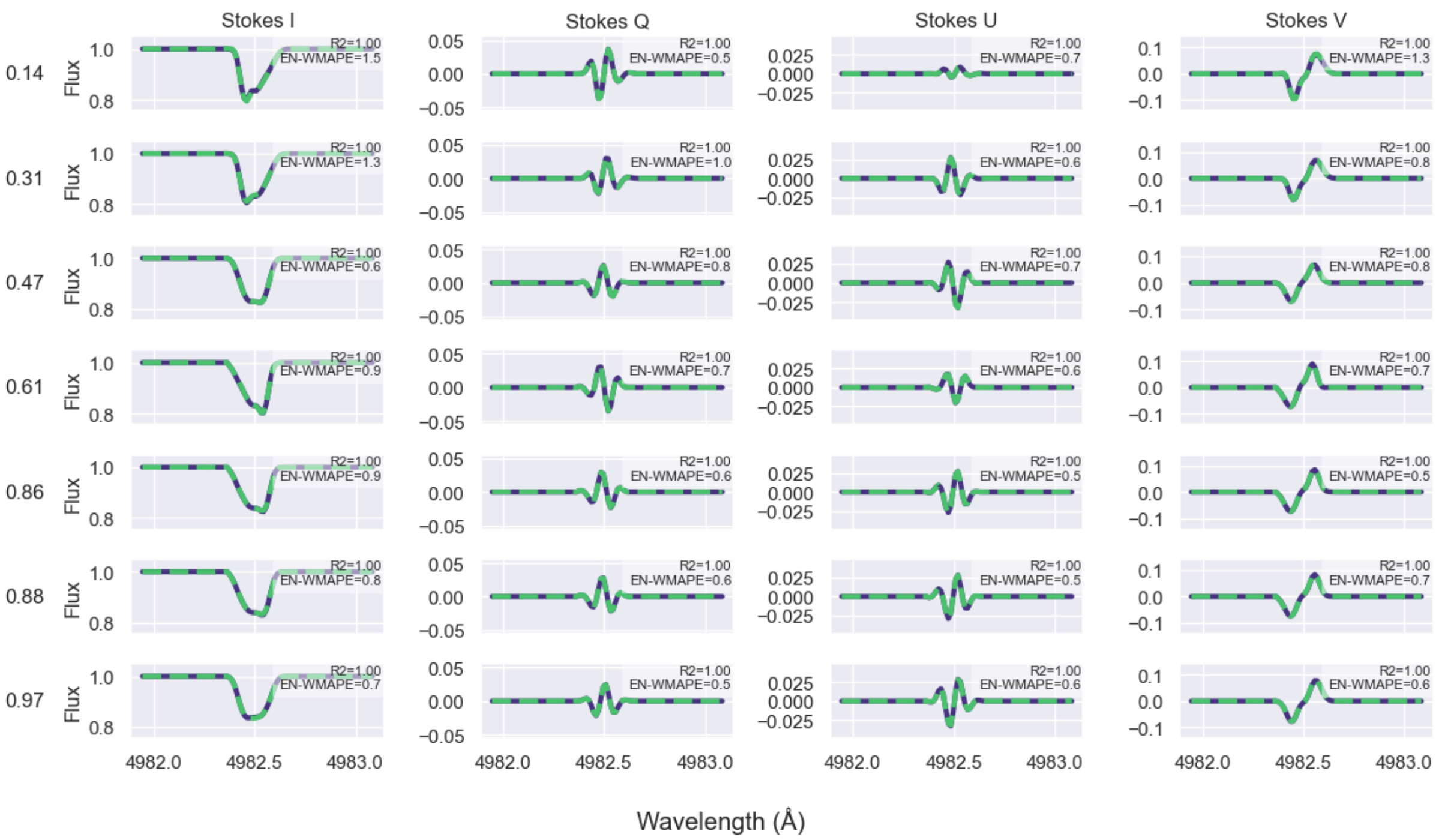

**Figure 2.** Fit of Stokes profiles at 7 different rotation phases; in dark blue are shown the profiles synthesized with COSSAM and in green those found as solution synthesized with MAPnet. In the title, the magneto-atmospheric parameters used by COSSAM are indicated with the legend *True*, while those found as solution are indicated with the legend *Pred*. The units of each parameter are indicated in Table 1, excepting the dipole position $x_r$ wich is given in units of $R_\star$.

relative to the continuum level, the number of spectral lines within this interval exhibits a strong dependence on metallicity, varying from as few as 8 lines at [M/H] = -2.0 dex to as many as 255 lines at [M/H] = +0.5 dex. As in previous section, line lists for each atmospheric model were retrieved from the VALD database (Ryabchikova et al. 2015).

We used the COSSAM code to establish the Stokes profiles in the mentioned spectra range of 100 Å, considering the same range of free parameters as before, i.e. those given in Table 1.

The are two main methods used to perform the line addition, namely, Least-Square-Deconvolution (LSD) and Singular Value Decomposition (SVD). The former is the most widely used (Donati et al. 1997; Kochukhov et al. 2010; Ramírez Vélez 2020), however in this approach the depth of each line must be know a priori to establish the mean profile. This is prohibitive in our case, since the depth of each line varies as function of metallicity, temperature and gravity. In other words, in the LSD approach these three stellar properties are fixed. On the contrary, the SVD approach does not rely in the prior knowledge of the line depths, i.e., [M/H], $T_{eff}$ and log g can be considered as free parameters. We therefore adopted the SVD strategy to obtain the multi-line profiles (Martínez González et al. 2008; Carroll et al. 2012).

To maintain consistency with the tests described in the previous section, it is first required that MAPNet demonstrates the ability to replicate multi-line Stokes profiles. It is important to clarify that MAPNet does not synthesize Stokes profiles across the full 100 Å, spectral range; instead, it is trained on mean-line profiles. Consequently, the network generates multi-line profiles as if they were produced by addition of individual lines within the specified spectral range.

The training procedure is based on the transfer learning framework also employed in Paper I. The main property of this methodology is the initialization of a new ANN using a pre-trained “seed” network, enabling the model to address a closely related task (Stokes profiles synthesis in this case). This approach markedly decreases the volume of training data required, in contrast to conventional training from scratch (Weiss et al. 2016). In the present study, the seed ANN corresponds to the model introduced in the preceding subsection, whereas the target network is specifically developed to synthesize multi-line Stokes profiles.

In practice, the transfer learning approach involves fixing the weights of the neurons in the initial hidden layers and retraining the seed ANN such that only the last layers are updated. Through this procedure, the new network is able to exploit representations previously learned by the seed model, thereby reducing the size of the training dataset required.

The ideal training set size was determined through multiple tests (see Paper I for details). The relationship between ANN performance and database size is presented in Fig. 4.

We observe that, even when trained with the smallest dataset considered (25k instances), the model is capable of achieving satisfactory performance relative to the baseline ANN trained on 1.5 million synthetic samples (seed model). Although a decrease in synthesis accuracy is observed for the polarized Stokes profiles (Q, U, and V), this behavior is expected given the reduced size of the retraining dataset and the increased complexity of reproducing polarization sig-natures. Nevertheless, the overall performance remains sufficiently high for the ANN to be reliably employed in the inversion process.

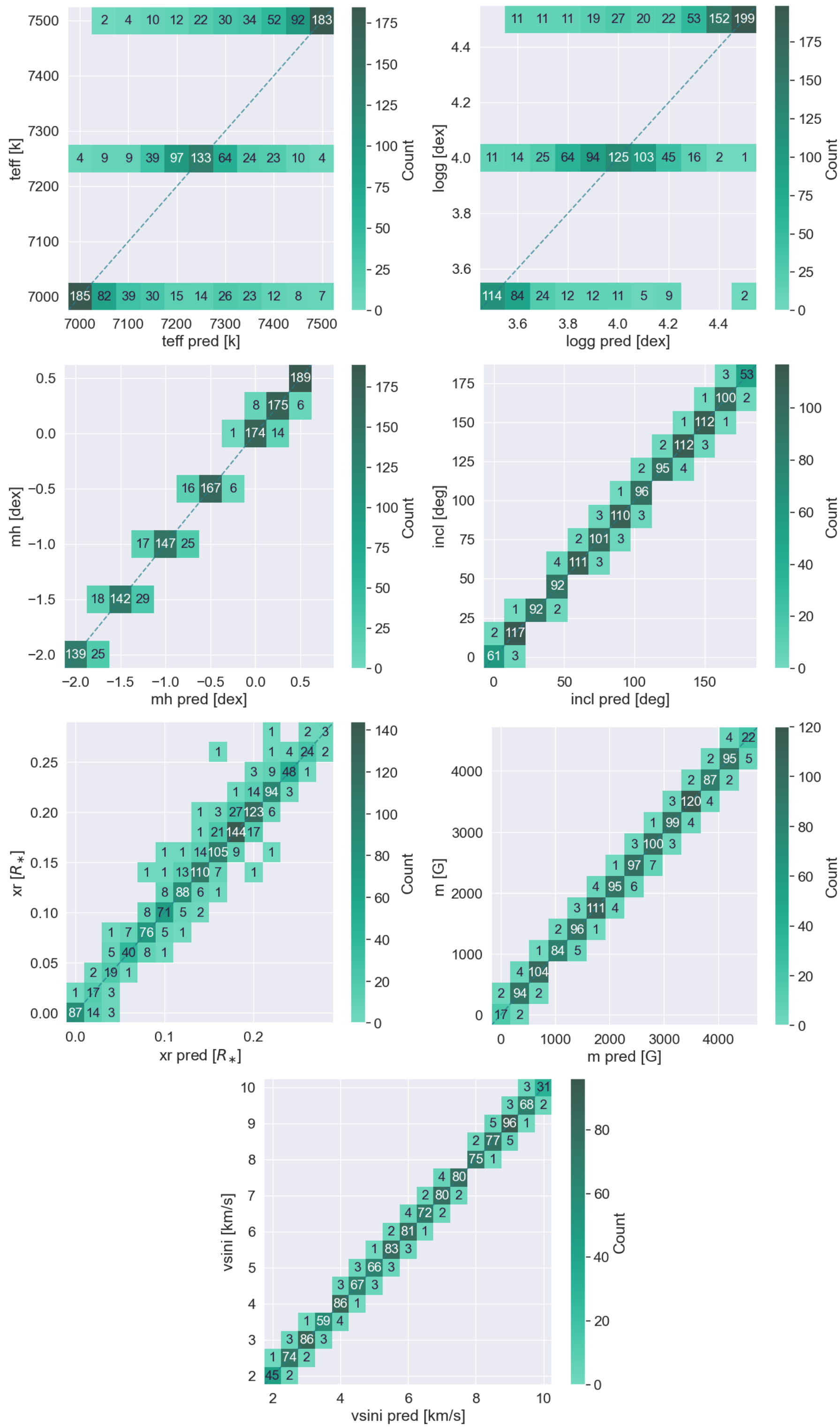

**Figure 3.** Two dimensional histograms of the inversion results of the full sample (1,298 cases) analyzing the Fe line at 4982.4 Å. In each bin are indicated the respective number of cases, also represented by the color bar. In the Y axis are the real values of the parameters and in the X axis the inferred solutions.

As expected, increasing the number of training instances leads to improved ANN performance. However, for multi-line profiles, it is generally preferable to rely on a relatively small training dataset, particularly when the profiles span spectral ranges of thousands of Å, since the time required to generate the training instances can become prohibitive. On this basis, we conclude that a dataset of 100k instances represents an appropriate compromise between computational cost and model performance.

**Table 3.** Same as Table 2 but considering multi-lines profiles instead of the single FeI line. The parameters with an $R^2$ equal or superior to 0.9 has been marked in bold.

| | $T_{eff}$ (K) | log g (dex) | [M/H] (dex) | Incl (deg) | $x_r$ ($R_\star$) | m (G) | vsini (km/s) |
|---|---|---|---|---|---|---|---|
| MAE | **26** | **0.08** | **0.03** | **1.0** | **0.007** | **31** | **0.04** |
| RMSE | **49** | **0.13** | **0.05** | **2.0** | **0.012** | **48** | **0.05** |
| $R^2$ | **0.942** | **0.900** | **0.997** | **0.998** | **0.970** | **0.998** | **0.999** |

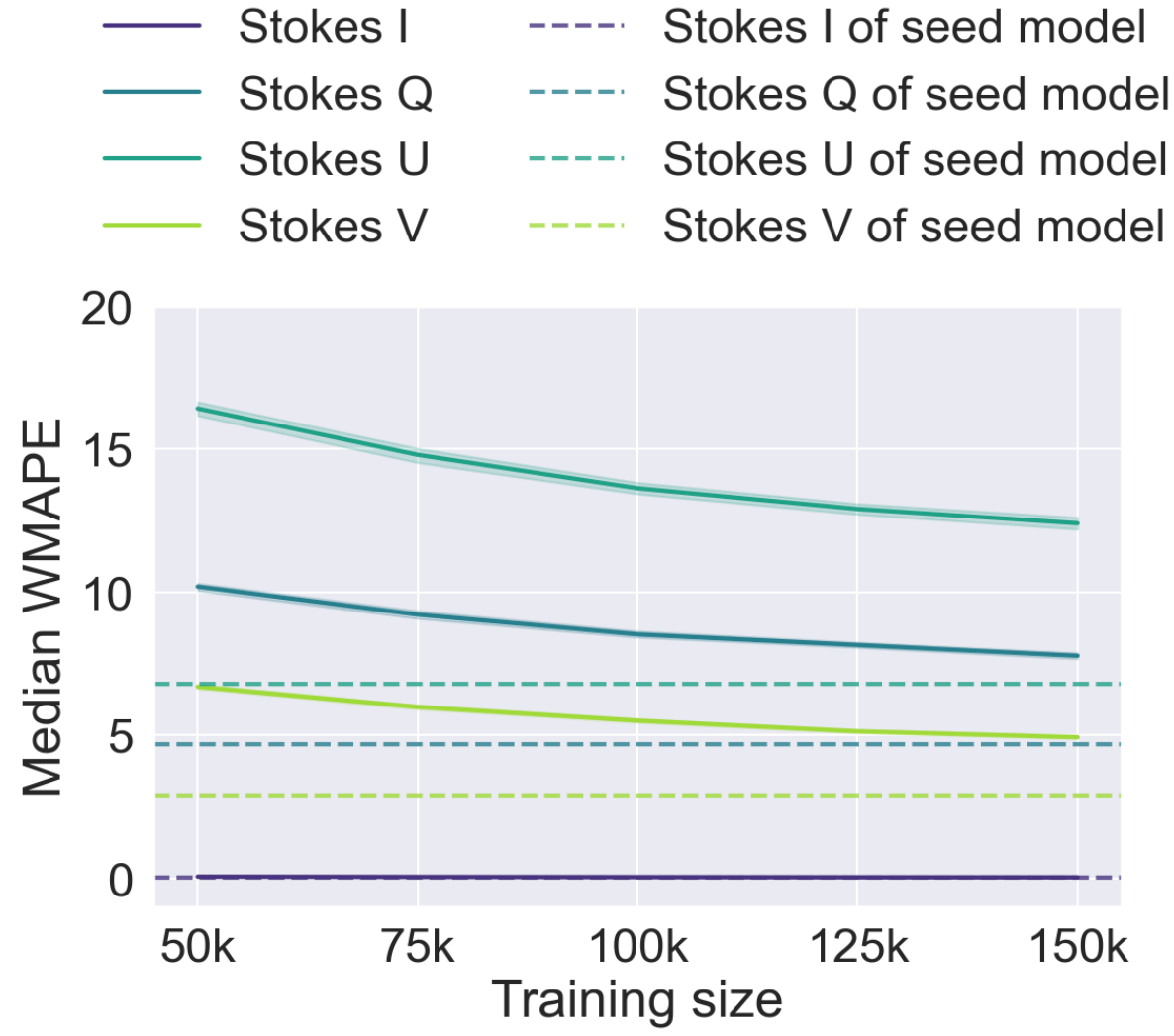

**Figure 4.** Performance of the ANN: weighted mean absolute percentage error (WMAPE) as function of the number of instances considered in the training database (solid lines); for completeness, we have included as reference the median WMAPE of the seed ANN trained with a database of 1.5 million instances in dashed lines.

The inversion of multi-line profiles follows the same methodology as before, namely, we considered 7 known phases $\phi_j$ and at each one with the COSSAM code we obtained the respective spectra (5000 to 5100 Å), to finally obtain for each phase a multi-line profile in each of the Stokes parameters. As before, the 11 magnetic and atmospheric parameters $\chi_i$ varied randomly in the ranges given in Table 1. We considered a sample of 1,008 "observed" stars.

The inversion results are resumed in Table 3, but consistently with the previous section, the fit of the Stokes profiles at percentile 75 and the 2D histograms are included in the appendix C, respectively Figs. C1 and C2.

Contrary to the case of the FeI line, now the temperature and the gravity, together with the rest of the parameters are well recovered ($R^2$ > 0.9 in all parameters). It has been accepted that the mean profiles encoded properly the information of the stellar magnetic field, and we are here proving that not only the magnetic but simultaneously also the atmospheric model can be inferred from multi-lines profiles with high accuracy.

## 4 DISCUSSION

Based on the good results obtained for the inversion of multi-line profiles, we are in possibility to inspect the impact of each free parameter in the inversion process. In particular, we concentrate in characterize the role of the rotation inclination angle $i$ respect to the LOS in the data analysis. To our knowledge, all stellar magnetic fields maps published so far from the inversion of a stellar spectropolarimetric data set in a given star, have assumed that the inclination angle is known prior to the data inversion process. In some studies, probably motivated by the pioneer work of Donati & Brown (1997), the inclination angle has been fixed to arbitrary angles, typically 60 degrees or close to (e.g. Bellotti et al. 2025), while in other studies the period, vsini and the radius of the star are used to determine the inclination angle (e.g. Abt et al. 1972; Masuda & Winn 2020). An alternative method –much less employed– to determine the angle $i$, is based in the analysis of broad band linear polarimetric data (Bagnulo et al. 1995).

Therefore, we evaluated the effect of fixing the rotation inclination angle. To introduce arbitrary errors, we intentionally set the inclination angle to random values between 0 and 180 degrees. Focusing on multi-line profiles, we repeated the precendet test with the sole modification of setting the inclination angle to a constant, arbitrary value. Additionally, we also considered two different observational scenarios, one in which the four Stokes parameters (I,Q,U,V) are at disposal for analysis and another one considering only Stokes (I,V) profiles.

Given that by far the most common observational case is to consider only Stokes (I,V), in the upper panels of Fig. 5 we show an example of a good fit of the Stokes profiles with a large difference (93 deg) in the inclination angle between the real and the assumed one. As consequence of this difference in the inclination angle, the inference of the magnetic moment is overestimated by 2,200 G (270%) and the position of the dipole is displaced to towards outside of the star by 0.1 $R_\star$. In the lower panels of Fig. 5, we show another fit to the Stokes profiles example but considering a small difference in the inclination angle of 17.5 deg. In this case, the position of the dipole remains well recovered, but the magnetic moment is now underestimated by 1246 G (443%). Therefore, it is concluded that when considering Stokes (I,V), to fix the inclination angle without certitude to a given value is critical when mapping the stellar magnetic fields.

For consistency, we show a similar fitting example considering the four Stokes profiles (I,Q,U,V) in appendix D, in Fig. D1, where for a difference of 26 degrees in the inclination angle $i$, the magnetic moment is now overestimated by 1,168 G (40%).

From these fitting examples, it can be concluded that the Stokes profiles may be satisfactorily reproduced for different assumed inclination angles, even when the recovered physical solutions are significantly different. This demonstrates the presence of strong degeneracies in the inversion problem, where visually good profile fits do not necessarily correspond to the correct stellar and magnetic configuration. Consequently, a visual inspection of the profile agreement alone is insufficient to validate the inferred solution. It is important to note that all experiments presented in this work were performed without the inclusion of observational noise; therefore, the presence of noise in real observations would likely increase the degeneracy of the solutions even further.

Tables 4 and 5 resume the inversions results using four and two Stokes parameters, respectively, and fixed inclination angle, over the full sample. It is clear that now the high accuracy obtained before is no longer preserved; only the metallicity and the vsini parameters (in the case of four Stokes profiles) preserves $R^2 > 0.9$.

Moreover, the most affected parameter corresponds to the position of the magnetic dipole inside of the star, despite if are considered two or four Stokes parameters ($R^2 < 0.12$ in both cases). The magnetic moment seems to be more resilient with $R^2 \sim 0.8$ in both cases, however please note that typical errors increases by one order of magnitude from RMSE less than 50 G (Table 3) to more than 500 G (Tables 4 and 5).

**Table 4.** Same as Table 2 but considering multi-lines profiles, assuming fixed random inclination angles, and using Stokes (I,Q,U,V)

| | $T_{eff}$ (K) | log g (dex) | [M/H] (dex) | Incl (deg) | $x_r$ ($R_\star$) | m (G) | vsini (km/s) |
|---|---|---|---|---|---|---|---|
| MAE | 56 | 0.13 | **0.07** | 60 | 0.049 | 368 | 0.43 |
| RMSE | 110 | 0.23 | **0.10** | 71 | 0.068 | 575 | 1.00 |
| $R^2$ | 0.710 | 0.659 | **0.986** | -0.91 | 0.117 | 0.793 | 0.801 |

**Table 5.** Same as Table 2 but considering multi-lines profiles, assuming fixed random inclination angles, and using Stokes (I,V)

| | $T_{eff}$ (K) | log g (dex) | [M/H] (dex) | Incl (deg) | $x_r$ ($R_\star$) | m (G) | vsini (km/s) |
|---|---|---|---|---|---|---|---|
| MAE | 48 | 0.12 | **0.06** | 60 | 0.050 | 340 | 0.38 |
| RMSE | 92 | 0.21 | **0.09** | 71 | 0.069 | 522 | 0.79 |
| $R^2$ | 0.798 | 0.722 | **0.989** | -0.91 | 0.081 | 0.829 | 0.877 |

Finally, to facilitate a direct comparison between the different inversion configuration tests presented so far (considering two/four Stokes parameters and fixing or not the inclination angle i), Fig. 6 presents the $R^2$ scores obtained for the inference of each physical parameter. The results show that the metallicity is recovered with nearly identical accuracy across all configurations, indicating that its estimation within the spectral interval tested relatively is insensitive to whether the inclination angle is treated as a free parameter or fixed during the inversion process. Similarly, parameters such as $T_{eff}$, log g, the dipolar magnetic field strength, and v sini have moderate degradation.

In contrast, the recovery of the dipole position parameter deteriorates dramatically. With the inclination angle fixed, the inversion is no longer capable of recovering the dipole position, as reflected by the near-zero $R^2$ values obtained. These results indicate that, within the explored parameter framework, fixing the inclination angle erroneously during the inversion process can introduce significant degeneracies in the recovered solutions.

## 5 CONCLUSIONS

In this work, we presented MAPNet, a methodology for the analysis of spectropolarimetric multi-line profiles that combines a neural-network-based synthesis model with a population-based optimization algorithm to simultaneously infer stellar atmospheric and magnetic parameters. One of the key advantages of MAPNet is its high synthesis speed, which is particularly beneficial for the case of multi-line profiles.

Across all experiments conducted in this work, a total of 6,628 inversion cases were performed. During these inversions, more than 3.3 billion candidate parameter combinations were explored, corresponding to over 23 billion synthesized sets of Stokes profiles. Despite this large computational workload, the complete set of experiments required less than 83 hours using a single GPU, demonstrating the computational efficiency and scalability of the proposed framework.

All tests were conducted in an hypothetical case without noise to facilitate the assessment of possible degeneracies in the solution among the free parameters itself, finding that when single line analysis is applied, the temperature and gravity showed certain inaccuracy in their recovery. However, when multi-line profiles were considered the proposed methodology is capable of recovering both atmospheric and magnetic parameters with high performance, achieving coefficients of determination above $R^2 > 0.9$ for all of the inferred parameters.

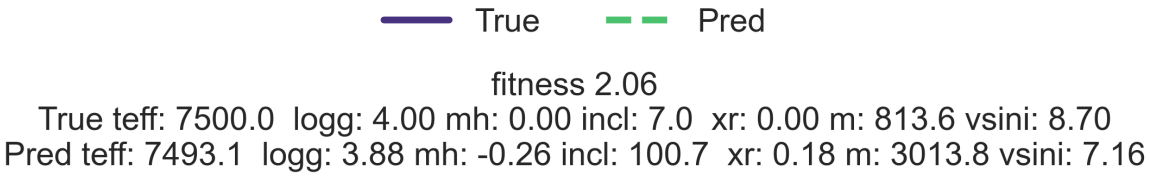

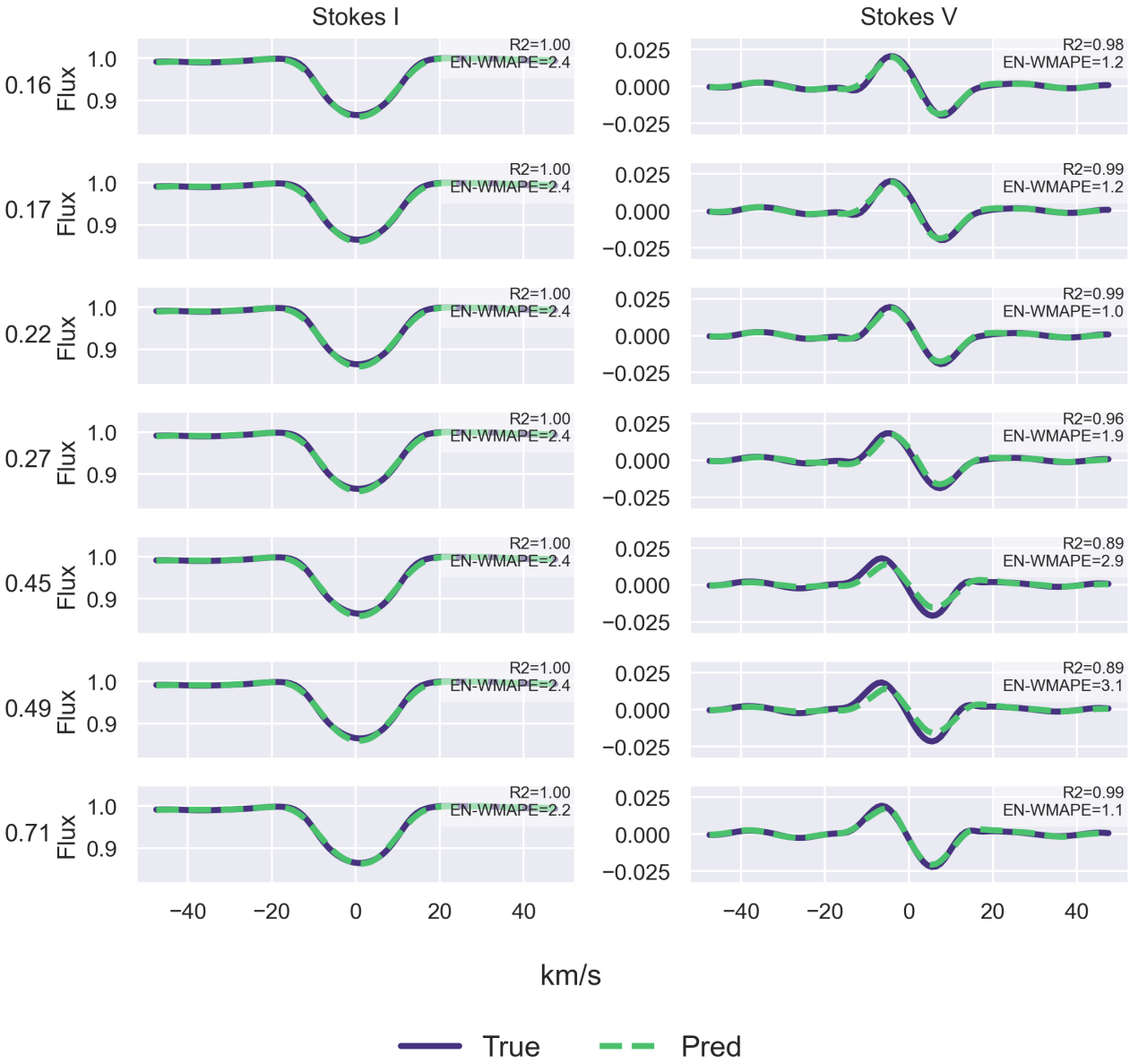

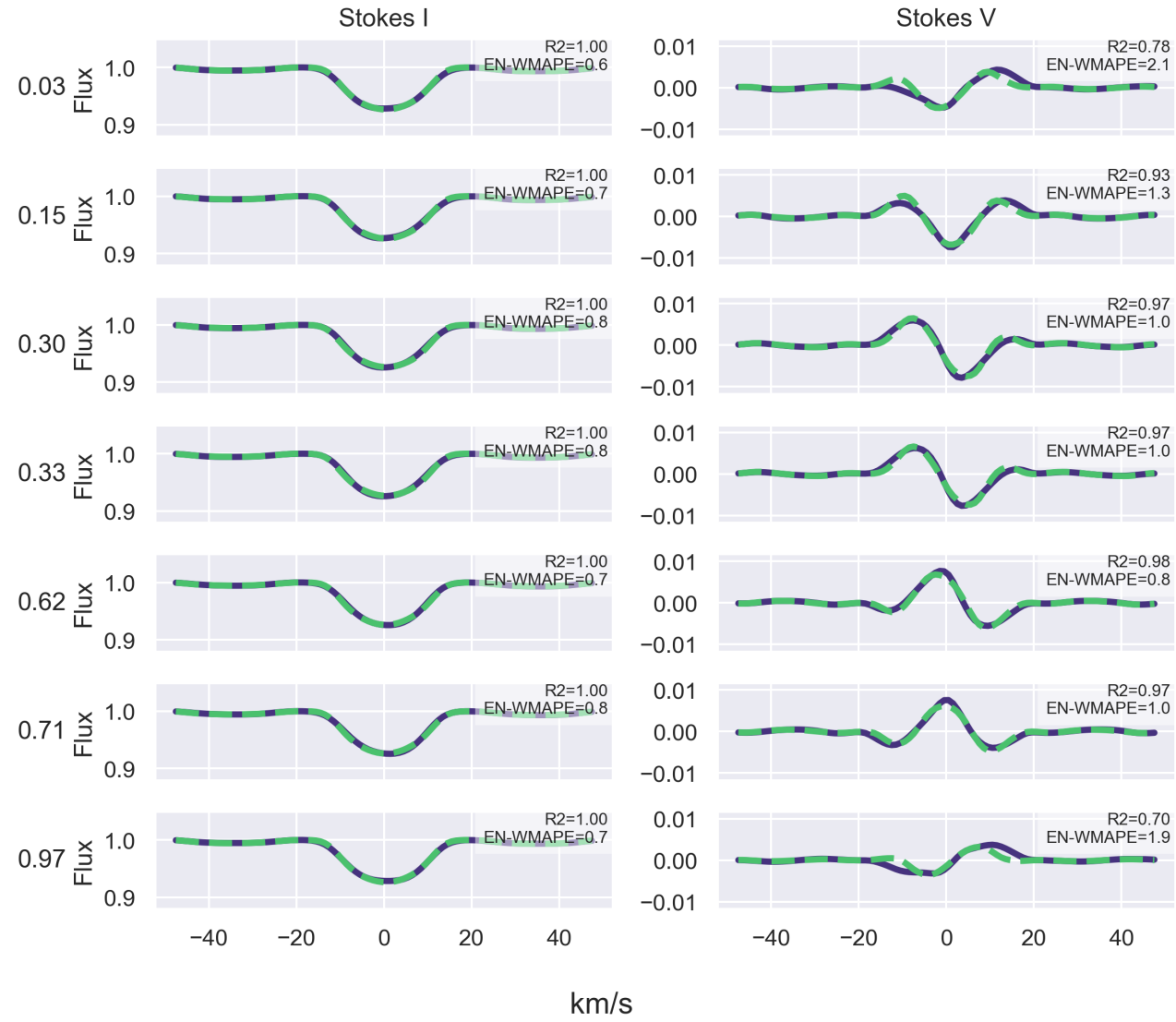

**Figure 5.** Two examples of multi-line profiles adjustment when the inversion assumes random inclination rotation angles and are considered only the Stokes (I,V); see text for details.

Notably, when atmospheric parameters are treated as free, the SVD approach is preferred over the LSD one to perform the line addition. This preference is because in practice SVD does not demand prior knowledge of the atmospheric parameters, contrary to LSD where the required line depths heavily depend on the free parameters themselves (temperature, gravity, and metallicity).

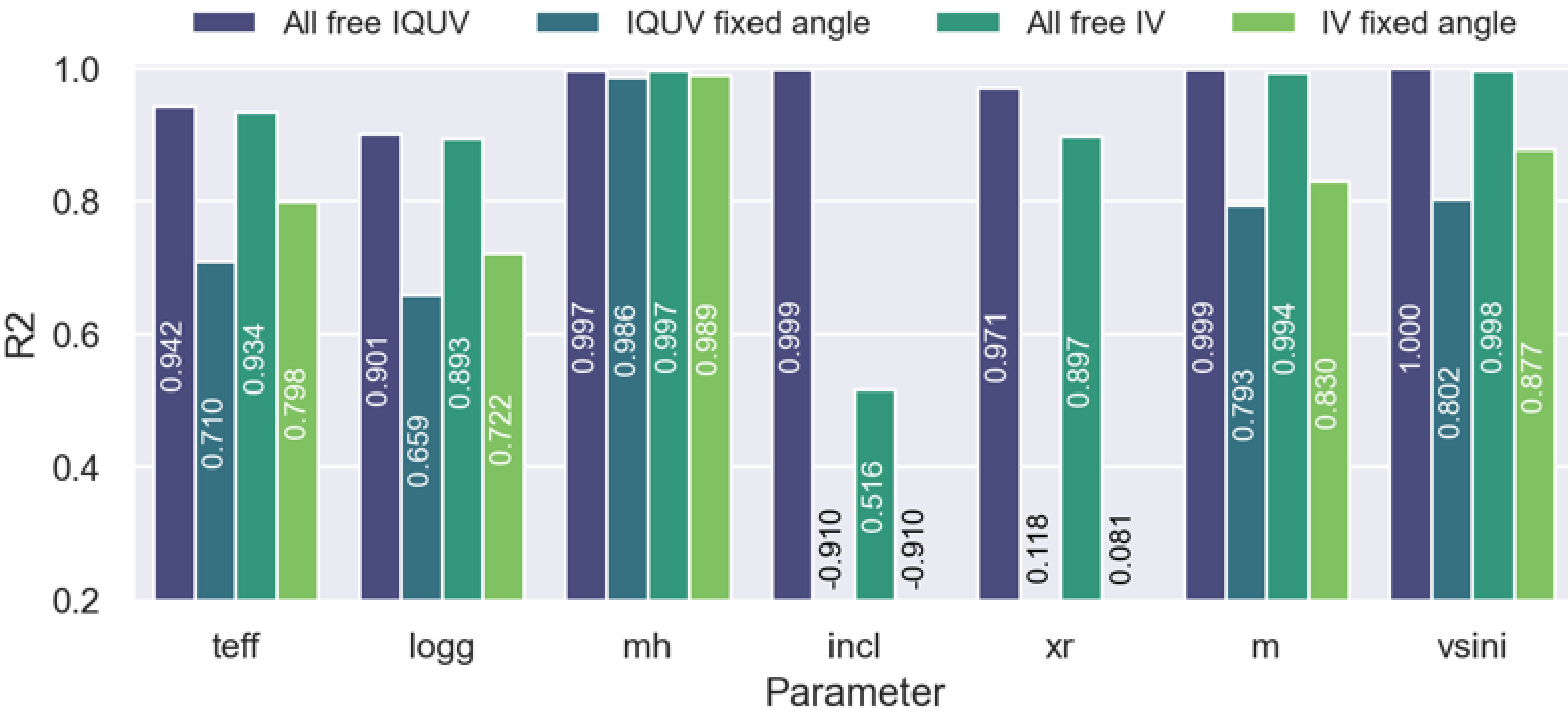

**Figure 6.** Comparison of the $R^2$ scores obtained for each inferred parameter under the different inversion configurations: using the full Stokes profiles (IQUV) or only Stokes IV, and considering either a free or fixed inclination angle during the inversion process.

Our results highlight the importance of simultaneously inferring both magnetic and atmospheric parameters during the inversion process. Specifically, fixing the inclination angle introduces significant degeneracies into the recovered solutions, particularly affecting the magnetic dipole position and the magnetic moment. Surprisingly, remarkable fits to the Stokes profiles can be achieved (as shown in Figs. 5 and D1) even when the inclination angle is fixed to an incorrect value.

Ultimately, the assumed inclination angle of the stellar rotation axis plays a crucial role; if this parameter is misspecified, then the stellar magnetic properties are no longer inferred properly.

While this work focuses on a de-centered dipolar geometry, expanding the framework to complex poloidal and toroidal magnetic fields necessitates higher-order $(l, m)$ spherical harmonics, thereby increasing the number of free parameters. Given our findings for the simplest configuration, misestimations in the assumed inclination angle will most likely lead to degeneracies among the $(l, m)$ coefficients—a formal characterization of which is still pending. Additionally, the observational sampling rate within a single stellar rotation period represents a critical factor to investigate, as phase coverage inherently limits the data inversion solution and could impact the results differently depending on magnetic complexity.

We plan to apply MAPNet to real data in a forthcoming study, and for that proper noisy tests have to be considered in order to determine the level of confidence expected in the results, particularly in the characterization of the magnetic properties.

## ACKNOWLEDGEMENTS

The authors thanks to Martin Stift to provide openly the use of the COSSAM code. JCRV acknowledges to the UNAM PAPIIT project 118023. R. L-V acknowledges support from Secretaría de Ciencia, Humanidades, Tecnología e Inovacción (SECIHTI) through a postdoctoral fellowship within the program “Estancias posdoctorales por México”

## DATA AVAILABILITY

There are no new data associated with this article.

## REFERENCES

Abt H. A., Chaffee F. H., Suffolk G., 1972, ApJ, 175, 779
Asensio Ramos A., Trujillo Bueno J., Landi Degl’Innocenti E., 2008, ApJ, 683, 542
Bagnulo S., Landi Degl’Innocenti E., Landolfi M., Leroy J. L., 1995, A&A, 295, 459
Bellotti S., et al., 2025, A&A, 693, A269
Brown S. F., Donati J.-F., Rees D. E., Semel M., 1991, A&A, 250, 463
Carroll T. A., Strassmeier K. G., 2014, A&A, 563, A56
Carroll T. A., Kopf M., Ilyin I., Strassmeier K. G., 2007, Astronomische Nachrichten, 328, 1043
Carroll T. A., Strassmeier K. G., Rice J. B., Künstler A., 2012, A&A, 548, A95
Castelli F., Kurucz R. L., 2003, in Piskunov N., Weiss W. W., Gray D. F., eds, IAU Symposium Vol. 210, Modelling of Stellar Atmospheres. p. A20 (arXiv:astro-ph/0405087), doi:10.48550/arXiv.astro-ph/0405087
Donati J.-F., Brown S. F., 1997, A&A, 326, 1135
Donati J.-F., Semel M., Carter B. D., Rees D. E., Collier Cameron A., 1997, MNRAS, 291, 658
Donati J.-F., et al., 2006, MNRAS, 370, 629
Finociety B., Donati J.-F., 2022, MNRAS, 516, 5887
Gutteridge C., Neiner C., Catala C., Folsom C. P., 2026, A&A, 709, A43
Jones D., Perttunen C., Stuckman B., 1993, , 79, 157
Kennedy J., 1999, in Proceedings of the 1999 Congress on Evolutionary Computation-CEC99 (Cat. No. 99TH8406). pp 1931–1938 Vol. 3
Kennedy J., Eberhart R., 1995, in Proceedings of ICNN’95 - International Conference on Neural Networks. pp 1942–1948 vol.4
Kochukhov O., 2017, A&A, 597, A58
Kochukhov O., Piskunov N., 2002, A&A, 388, 868
Kochukhov O., Makaganiuk V., Piskunov N., 2010, A&A, 524, A5
Kochukhov O., Mantere M. J., Hackman T., Ilyin I., 2013, A&A, 550, A84

Martínez González M. J., Asensio Ramos A., Carroll T. A., Kopf M., Ramírez Vélez J. C., Semel M., 2008, A&A, 486, 637
Masuda K., Winn J. N., 2020, AJ, 159, 81
Pedregosa F., et al., 2011, Journal of Machine Learning Research, 12, 2825
Piskunov N., Kochukhov O., 2002, A&A, 381, 736
Piskunov N., et al., 2011, The Messenger, 143, 7
Press W. H., Flannery B. P., Teukolsky S. A., Vetterling W. T., 1986, Numerical recipes: The art of scientific computing
Ramírez Vélez J. C., 2020, MNRAS, 493, 1130
Raygoza-Romero J. M., Ramírez-Vélez J. C., Lopez-Nava I. H., 2025, MNRAS, 542, 2404
Ruiz Cobo B., del Toro Iniesta J. C., 1992, ApJ, 398, 375
Ruiz Cobo B., Quintero Noda C., Gafeira R., Uitenbroek H., Orozco Suárez D., Páez Mañá E., 2022, A&A, 660, A37
Ryabchikova T., Piskunov N., Kurucz R. L., Stempels H. C., Heiter U., Pakhomov Y., Barklem P. S., 2015, Phys. Scr., 90, 054005
Semel M., Li J., 1996, Sol. Phys., 164, 417
Socas-Navarro H., de la Cruz Rodríguez J., Asensio Ramos A., Trujillo Bueno J., Ruiz Cobo B., 2015, A&A, 577, A7
Stift M. J., 1975, MNRAS, 172, 133
Stift M. J., Leone F., 2026, ApJ, 1003, 111
Stift M. J., Leone F., Cowley C. R., 2012, MNRAS, 419, 2912
Weiss K., Khoshgoftaar T. M., Wang D., 2016, Journal of Big data, 3, 1
de la Cruz Rodríguez J., Leenaarts J., Danilovic S., Uitenbroek H., 2019, A&A, 623, A74

## APPENDIX A: ERRORS AS FUNCTION OF THE PROFILE AMPLITUDE

Figure A1 illustrates the relationship between the synthesis error, measured in terms of WMAPE, and the amplitude of the Stokes profiles for the 4982.4 Å line. For this analysis, the amplitude of a Stokes profile is defined as the difference between its maximum and minimum values.

A clear dependence of the synthesis error on the profile amplitude can be observed, particularly for the polarized Stokes parameters ($Q$, $U$, and $V$). Profiles with very low amplitudes exhibit larger synthesis errors and a wider dispersion in WMAPE values. As the amplitude increases, both the median error and its variability decrease considerably, indicating that stronger polarization signatures are synthesized more reliably by MAPNet. This behavior is expected, since weak polarization signals are intrinsically more difficult for the ANN to reproduce accurately due to their lower signal strength and reduced contrast relative to the continuum.

In contrast, Stokes $I$ exhibits a weaker dependence on profile amplitude, with relatively stable synthesis errors across amplitude ranges. This can be attributed to the fact that intensity profiles generally present stronger and less noisy signals than the polarized components, making them easier to reproduce consistently.

## APPENDIX B: INFLUENCE OF STOKES PROFILE AMPLITUDE ON INVERSION PERFORMANCE

In this section, we address the effect of the amplitude-dependent synthesis errors on the inversion process and propose a strategy to mitigate their impact. As discussed in Appendix A, the synthesis accuracy of MAPNet is not uniform across all Stokes profile amplitudes. In particular, polarization profiles with very low amplitudes tend to exhibit systematically larger reconstruction errors, which may introduce biases during the inversion.

Throughout this work, the weighted mean absolute percentage error (WMAPE) is employed to quantify the agreement between the Stokes profiles synthesized with the radiative transfer code COSSAM and those generated by MAPNet. While WMAPE provides a convenient normalized metric for comparing the four Stokes parameters, preliminary analyses revealed a significant dependence of this metric on profile amplitude, especially for Stokes $Q$, $U$, and $V$, whereas the dependence is considerably weaker for Stokes $I$ (see Fig. A1).

This behavior implies that low-amplitude polarization signatures naturally produce larger relative synthesis errors, even when the neural network reconstruction remains physically consistent. As a consequence, directly using a conventional WMAPE-based fitness function may overemphasize low-amplitude polarization profiles, as their systematically larger relative errors can contribute disproportionately to the optimization objective. This introduces an imbalance in the relative influence of the Stokes parameters during the inversion process.

To mitigate this effect, we define an *Expected-Error-Normalized Weighted Mean Absolute Percentage Error* (EN-WMAPE), in which the WMAPE of each Stokes profile is normalized by the empirically expected synthesis error of the ANN as a function of profile amplitude. This normalization compensates for the amplitude-dependent synthesis performance of MAPNet and prevents profiles with intrinsically different expected error levels from contributing disproportionately to the optimization process.

The inversion fitness function is defined as

$$\text{EN-WMAPE} = \frac{1}{N_\phi N_S} \sum_{i=1}^{N_\phi} \sum_{s=1}^{N_S} \frac{\text{WMAPE}\left(\mathbf{y}_{i,s}^{\text{obs}}, \mathbf{y}_{i,s}^{\text{syn}}\right)}{\widehat{\text{WMAPE}}_s(A_{i,s})} \tag{B1}$$

where the WMAPE between an observed and synthesized profile is defined as

$$\text{WMAPE}\left(\mathbf{y}^{\text{obs}}, \mathbf{y}^{\text{syn}}\right) = \frac{\sum_{k=1}^{N_\lambda} \left| y_k^{\text{obs}} - y_k^{\text{syn}} \right|}{\sum_{k=1}^{N_\lambda} \left| y_k^{\text{obs}} \right|} \tag{B2}$$

The expected synthesis error is modeled independently for each Stokes parameter as a function of profile amplitude using empirical fits derived from neural network validation experiments. Since the amplitude–error relation differs among Stokes parameters, different functional forms were adopted according to the best empirical fit. For Stokes $I$, the expected WMAPE is modeled using a fourth-order polynomial,

$$\widehat{\text{WMAPE}}_I(A) = a_I A^4 + b_I A^3 + c_I A^2 + d_I A + e_I \tag{B3}$$

whereas for Stokes $Q$, $U$, and $V$, a double-exponential function is employed,

$$\widehat{\text{WMAPE}}_s(A) = c_s + a_{1,s} \exp(-b_{1,s} A) + a_{2,s} \exp(b_{2,s} A), \in \{Q, U, V\} \tag{B4}$$

where $A$ represents the amplitude of the Stokes profile, $N_\phi$ denotes the number of rotational phases, and $N_S$ the number of Stokes parameters considered in the inversion. The vectors $\mathbf{y}_{i,s}^{\text{obs}}$ and $\mathbf{y}_{i,s}^{\text{syn}}$ correspond to the observed and synthesized Stokes profiles, respectively, for rotational phase $i$ and Stokes parameter $s$. The coefficients of the fitting functions were estimated independently for each Stokes parameter from test data of the ANN.

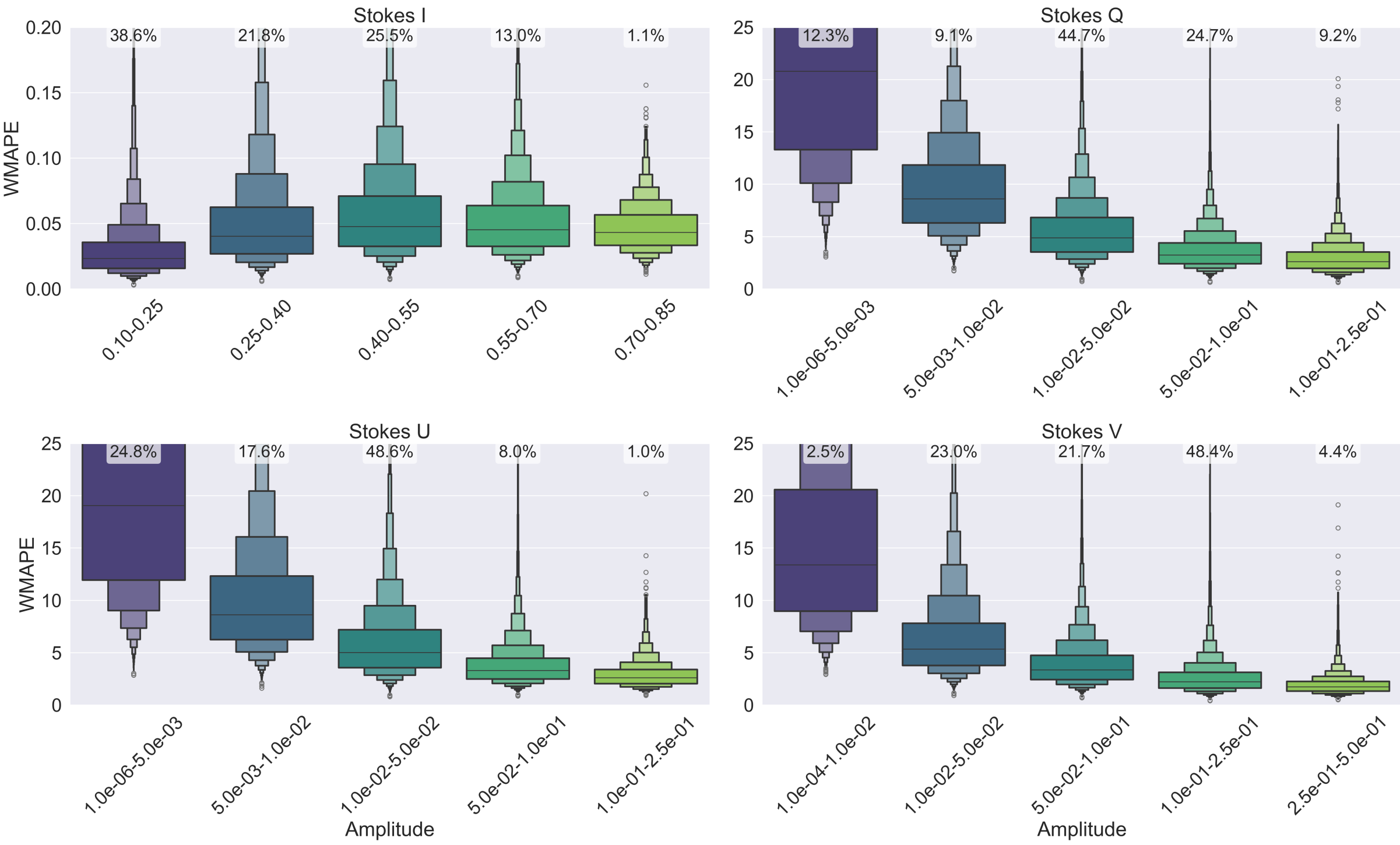

**Figure A1.** MAPNet error in the synthesis profile versus Stokes profile amplitude for the 4982.4 line. In top of each bin are indicated the percentage of the sample contained in that range of amplitudes.

## APPENDIX C: MULTI-LINE PROFILES ALL FREE PARAMETERS

In this appendix, we present the inversion performance obtained when all atmospheric and magnetic parameters are simultaneously inferred while considering the full Stokes vector (I,Q,U,V). Figure C1 shown an inversion case that is the 75th percentile based on fitness, and Figure C2 shown two-dimensional histograms comparing the inferred parameters against their corresponding ground-truth values.

## APPENDIX D: MULTI-LINE PROFILES FIXED INCLINATION ANGLE

This paper has been typeset from a TEX/LATEX file prepared by the author.

**Figure C1.** Same as Fig. Fig. 2 but considering multi-line profiles instead of the single FeI line at 4982.4 Å.

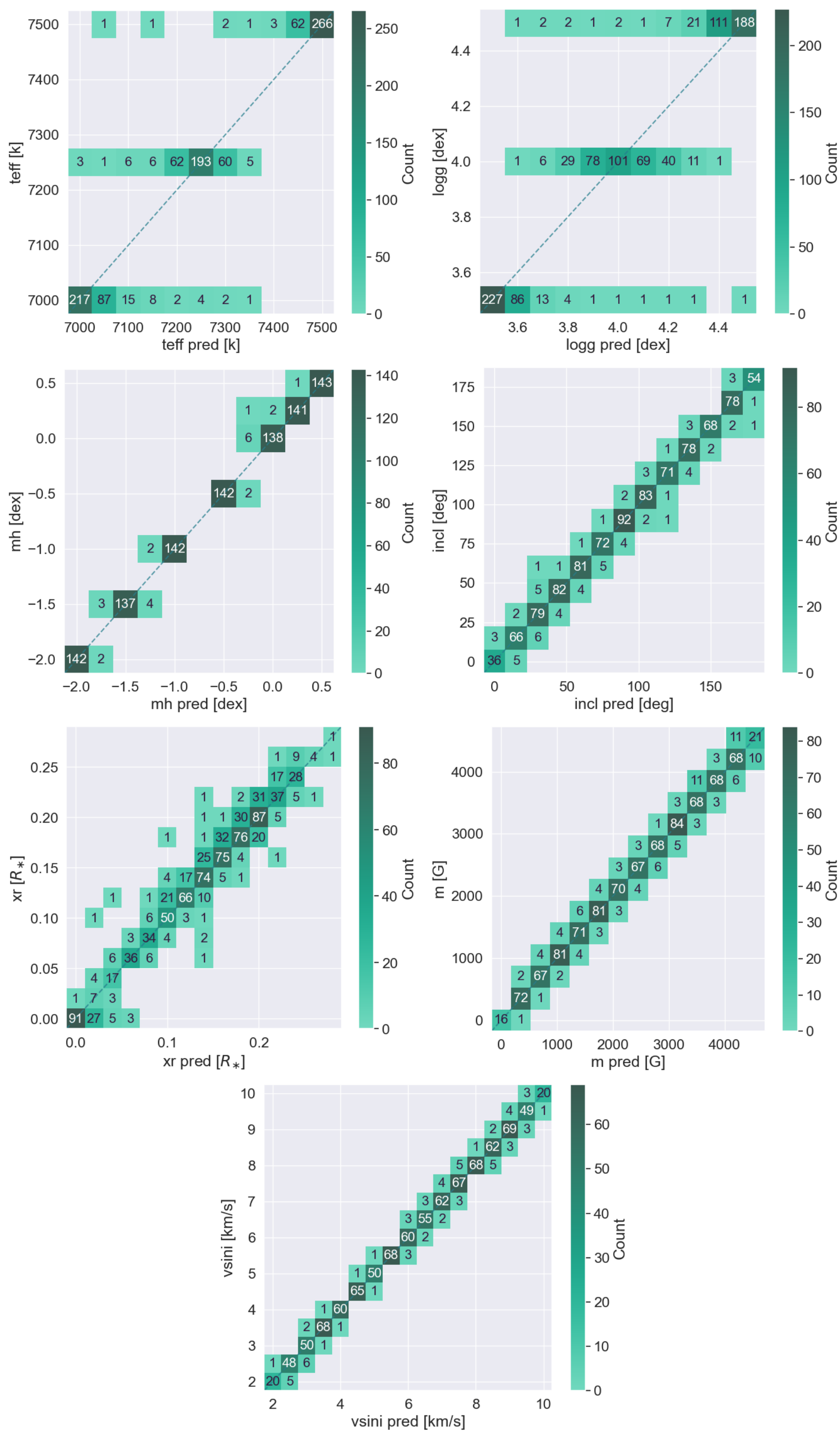

**Figure C2.** Same as Fig. 3 but considering inversion of multi-line profiles instead of the single FeI line at 4982.4 Å.

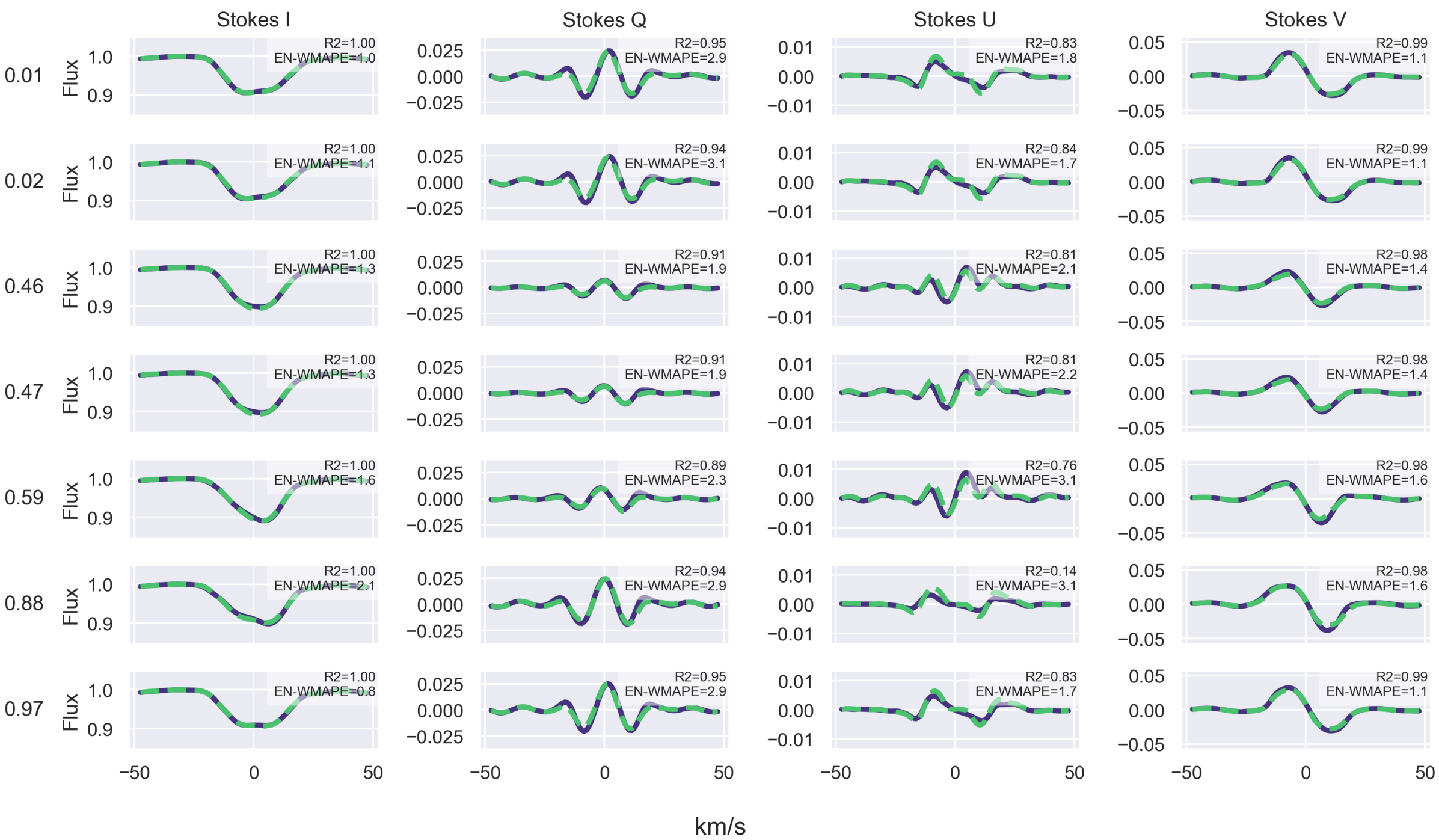

**Figure D1.** Same as Fig. 5 but considering the four Stokes profiles.